# On the Theoretical Status of Deep Inelastic Scattering

J. Blümlein

*DESY–Zeuthen,
Platanenallee 6, D–15735 Zeuthen, Germany*

**Abstract**

The theoretical status of perturbative QED and QCD corrections to deep inelastic scattering is reviewed.

# ON THE THEORETICAL STATUS OF
# DEEP INELASTIC SCATTERING

JOHANNES BLÜMLEIN

*DESY–Zeuthen, Platanenallee 6,*

*D–15735 Zeuthen, Germany*

e-mail: H1KBLU@dsyibm.desy.de

ABSTRACT

The theoretical status of perturbative QED and QCD corrections to deep inelastic scattering is reviewed.

1. Introduction
2. QED radiative corrections to DIS
3. The running coupling constant
4. The evolution equation
4.1. Splitting functions
4.2. Coefficient functions
4.3. $O(\alpha_s^3)$ corrections
5. Resummation of small $x$ contributions
5.1. Singlet terms
5.2. Non-singlet terms
6. Heavy flavour contributions to structure functions
7. $J/\psi$ production
8. QCD corrections to polarized structure functions
9. Open problems

## 1. Introduction

Deep inelastic lepton–hadron scattering provides one of the cleanest ways to investigate the nucleon structure at short distances. Since both charged leptons $(e^\pm, \mu^\pm)$ and neutrinos $(\nu, \overline{\nu})$ may be used as probes in neutral and charged current deep inelastic scattering off protons or isoscalar targets a variety of scattering cross sections can be measured containing different flavour combinations of quarks. Furthermore, the spin structure of nucleons can be investigated using both polarized leptons and targets.

The basic diagram describing the process is shown in figure 1. The neutral and charged current scattering cross sections are given by

$$\frac{d^2\sigma_{NC}^{lN}}{dxdQ^2} = P_{l,B}(Q^2)\frac{M_N sy}{x(s-M_N^2)^2}L_B^{\mu\nu}W_{\mu\nu}^B \tag{1}$$

in the Born approximation. The leptonic and hadronic tensors $L_B^{\mu\nu}, W_{\mu\nu}^B$ depend on the quantum numbers of the lepton and hadron, respectively, and those of the exchanged boson [1]. Here, $x$ and $y$ denote the Bjorken variables, $Q^2 = (l - l')^2$, and $s = (l + P)^2$. The factor $P_{l,B}(Q^2)$ collects couplings and propagator terms, e.g. $P_{l^\pm,\gamma}(Q^2) = 2\pi\alpha^2/Q^4$, $P_{\nu,W^+}(Q^2) = G_F^2 M_W^4/(4\pi(Q^2 + M_W^2)^2)$, with $\alpha$ being the fine structure constant, $G_F$ the Fermi constant, and $M_W$ the $W$ boson mass.

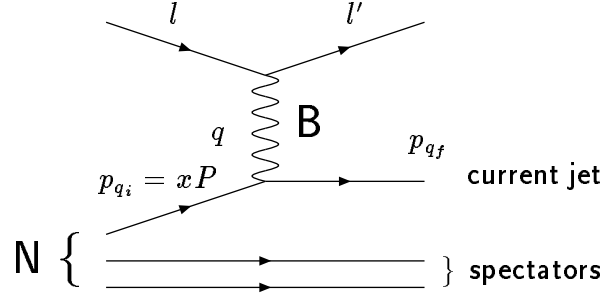

Figure 1: Diagram describing deep inelastic $lN$ scattering, with $B = \gamma, Z, W^\pm$, $N = p, n$, and $l = l^\pm, \nu, \overline{\nu}$.

The hadronic tensor may be represented by the structure functions describing the scattering process. In the parton model the structure functions are expressed in terms of parton densities, $q_i(x, Q^2)$. For $l^\pm N$ scattering the structure functions are thus given by:

$$\begin{aligned}
F_2(x, Q^2) &= x\sum_q e_q^2 [q(x, Q^2) + \overline{q}(x, Q^2)] & |\gamma|^2 \\
G_2(x, Q^2) &= 2x\sum_q e_q v_q [q(x, Q^2) + \overline{q}(x, Q^2)] & |\gamma Z| \\
H_2(x, Q^2) &= x\sum_q (v_q^2 + a_q^2)[q(x, Q^2) + \overline{q}(x, Q^2)] & |Z|^2 \\
xG_3(x, Q^2) &= 2x\sum_q e_q a_q [q(x, Q^2) - \overline{q}(x, Q^2)] & |\gamma Z| \\
xH_3(x, Q^2) &= x\sum_q v_q a_q [q(x, Q^2) - \overline{q}(x, Q^2)] & |Z|^2 \\
W_2^+(x, Q^2) &= 2x\sum_i [d_i(x, Q^2) + \overline{u}_i(x, Q^2)] & |W^+|^2 \\
W_2^-(x, Q^2) &= 2x\sum_i [u_i(x, Q^2) + \overline{d}_i(x, Q^2)] & |W^-|^2 \\
xW_3^+(x, Q^2) &= 2x\sum_i [d_i(x, Q^2) - \overline{u}_i(x, Q^2)] & |W^+|^2 \\
xW_3^-(x, Q^2) &= 2x\sum_i [u_i(x, Q^2) - \overline{d}_i(x, Q^2)] & |W^-|^2
\end{aligned} \quad (2)$$

where $e_q, v_q$, and $a_q$ denote the charge, vector-, and axialvector coupling constants of the quarks. The $\nu(\overline{\nu})N$ scattering processes are described by six further structure functions $W_2^{\nu,\pm}$, $xW_3^{\nu,\pm}$, $F_{2Z}$, and $xF_{3Z}$, in lowest order. In $\mathcal{O}(\alpha_s)$ also the longitudinal structure functions

$$\mathcal{S}_L(x, Q^2) = \mathcal{S}_2(x, Q^2) - 2x\mathcal{S}_1(x, Q^2), \quad (3)$$

with $\mathcal{S} \equiv F, G, H, W^{\pm}$, and $F_Z$, contribute to the scattering cross sections.

In the case of polarized lepton–polarized hadron scattering a similar amount of structure functions occurs [2]. In the kinematical range of the present experiments the scattering cross section is determined by the $|\gamma|^2$ term, however, and in this approximation only the structure functions $g_1(x, Q^2)$ and $g_2(x, Q^2)$ contribute.

Not all of the structure functions mentioned above can be determined at a sufficient accuracy combining different cross section measurements [3]. The structure functions $F_2^{l^{\pm} p(d)}(x, Q^2)$, and the combinations $W_2^{\nu N}(x, Q^2) \equiv \frac{1}{2}\left(W_2^+ + W_2^-\right)$, and $xW_3^{\nu N}(x, Q^2) \equiv \frac{1}{2}\left(xW_3^+ + xW_3^-\right)$ can be measured precisely in two variables $(x, Q^2)$ in a wide kinematical range. For them QCD analyses may be performed to determine the QCD parameter $\Lambda$ and constraints on the gluon density.

The measurement of the longitudinal structure functions $F_L^{l^{\pm} p(d)}(x, Q^2)$ is more difficult and requires a variation of the CMS energy keeping $x$ and $Q^2$ fixed. A precise determination of this structure function is of special importance since it is directly related to the gluon density, see sect. 4.2. For other structure functions as $xG_3^{l^{\pm} p(d)}$, or $F_{2Z}$, only the $x$-shape may be determined since they are measured from cross section differences, or, the reconstruction of the kinematical variables turns out to be difficult.

The determination of the QCD and QED corrections to the different deep inelastic scattering processes is of great importance for the quantitative understanding of the nucleon structure. In the present paper we give a survey on the status of perturbative QED and QCD radiative corrections to deep inelastic scattering processes. We also include a discussion of the status of the determination of $\alpha_s$, and of the QCD corrections to some exclusive processes, as the heavy flavour structure functions and $J/\psi$ production in $lN$ scattering, which play an important role for the determination of the gluon density of the proton.

## 2. QED Radiative Corrections to DIS

The QED radiative corrections to the deep inelastic scattering cross sections may become rather larger in some kinematical ranges. They have to be precisely known to unfold the neutral and charged current structure functions from the scattering cross sections. The first dedicated calculation of the radiative corrections to deep inelastic $eN$ scattering was performed by Mo and Tsai [4] and used in the analysis of the SLAC experiments. Later calculations were performed in refs. [5-6] for $l^{\pm}N$ scattering. The detailed knowledge of QED and electroweak radiative corrections was of special importance also for the measurements of the electroweak parameters in deep inelastic $\nu(\overline{\nu})N$ scattering [7-14]. With the advent of HERA the radiative corrections were partly recalculated and dedicated calculations for deep inelastic neutral and charged current $e^{\pm}p$ scattering were carried out by different groups using different techniques [15-30]. These approaches include both semi-analytical calculations [15-27] and calculations based on Monte Carlo techniques [28-30].

Dominant contributions to the QED radiative corrections may be obtained using leading log (LLA) techniques [16-24]. This approach, which is based on the factorization of (collinear) fermion mass singularities, allows to determine the terms $\propto \alpha \ln(Q^2/m_f^2)$ in a straightforward way for different settings of the measured kinematical variables. Also higher order terms were calculated within this approach [21,24]. The LLA QED radiative corrections may be described by

$$\frac{d^2\sigma^{LLA}}{dxdy} = \frac{d^2\sigma^0}{dxdy} + \frac{d^2\sigma^{l,1loop}}{dxdy} + \frac{d^2\sigma^C}{dxdy} + \frac{d^2\sigma^{l,2loop}}{dxdy} + \frac{d^2\sigma^{l,>2,soft}}{dxdy} + \frac{d^2\sigma^{l,e^-\to e^+}}{dxdy} + \ldots \quad (4)$$

Here, $l$ labels the type of bremsstarhlung, which is in lowest order either initial or final state radiation. In higher orders also products of initial and final state radiation terms contribute. In many situations final state radiation does not occur, or can be delt with in a cumulative way due to the calorimetric measurement of the scattered electron. Then all terms refer to initial state radiation only. The label $C$ denotes the Compton-contribution,[a] i.e. the collinear term formed by low $Q^2$ radiation of the virtual photon from the initial state hadron or quark lines. The soft contributions, $> 2, soft$, can be exponentiated. Starting with $\mathcal{O}(\alpha^2)$ also terms due to $e^- \to e^+$ conversion are present in the leading logarithmic order.

The first order terms are described by:

$$\frac{d^2\sigma^{ini(fin),1loop}}{dxdy} = \frac{\alpha}{2\pi}L_e \int_0^1 dz P_{ee}^{(1)} \left\{ \theta(z-z_0)\mathcal{J}(x,y,Q^2) \left.\frac{d^2\sigma^0}{dxdy}\right|_{x=\hat{x},y=\hat{y},S=\hat{S}} - \frac{d^2\sigma^0}{dxdy} \right\}, \quad (5)$$

where

$$P_{ee}^{(1)} = \frac{1+z^2}{1-z} \quad (6)$$

denotes the non-singlet QED splitting function of a massless fermion into a fermion. The scale of the correction is set by the logarithm

$$L_e = \ln\frac{Q^2}{m_e^2} - 1. \quad (7)$$

This notion reproduces the soft photon terms of complete calculations in leptonic variables (cf. e.g. [27]). The shifted variables $\hat{x}, \hat{y}$, and the Jacobian $\mathcal{J}$ depend on the choice of the outer kinematical variables (see [27] for a summary of these terms).

The Compton term is given by [17,18]

$$\frac{d^2\sigma^C}{dx_l dy_l} = \frac{\alpha^3}{x_l S}\left[1+(1-y_l)^2\right]\ln\left(\frac{Q_l^2}{M_N^2}\right) \int_{x_l}^1 \frac{dz}{z^2}\frac{z^2+(x_l-z)^2}{x_l(1-y_l)} \sum_f \left[q_f(z,Q_l^2)+\bar{q}_f(z,Q_l^2)\right] \quad (8)$$

---

[a]This term was already found in ref. [4].

for leptonic variables in LLA. A more refined expression was derived in ref. [22]. LLA second order corrections are easily obtained by convoluting with the leading order NS-splitting function. Although the Compton-type contribution counts to the radiative corrections to deep inelastic scattering in an inclusive description, its experimental signature is rather different compared to typical deep inelastic events, showing a photon–electron pair which is nearly balanced in $p_\perp$ and little hadronic activity only [22]. Such a signature can be easily tagged. Due to this one may even use these events to measure nucleon structure functions *both* at small $x$ and small $Q^2$.

The second order corrections $\mathcal{O}((\alpha L_e)^2)$ are:

$$\frac{d^2\sigma^{l,2loop}}{dxdy} = \left[\frac{\alpha}{2\pi}L_e\right]^2 \int_0^1 dz P_{ee}^{(2,1)}(z) \left\{\theta(z-z_0)\mathcal{J}(x,y,z)\frac{d^2\sigma^0}{dxdy}\bigg|_{x=\hat{x},y=\hat{y},S=\hat{S}} - \frac{d^2\sigma^0}{dxdy}\right\}$$

$$+ \left(\frac{\alpha}{2\pi}\right)^2 \int_{z_0}^1 dz \left\{L_e^2 P_{ee}^{(2,2)}(z) + L_e \sum_{f=l,q} \ln\frac{Q^2}{m_f^2} P_{ee,f}^{(2,3)}(z)\right\} \mathcal{J}(x,y,z)\frac{d^2\sigma^0}{dxdy}\bigg|_{x=\hat{x},y=\hat{y},S=\hat{S}}. \quad (9)$$

Here the different second order splitting functions are given by

$$P_{ee}^{(2,1)}(z) = \frac{1}{2}\left[P_{ee}^{(1)} \otimes P_{ee}^{(1)}\right](z)$$

$$= \frac{1+z^2}{1-z}\left[2\ln(1-z) - \ln z + \frac{3}{2}\right] + \frac{1}{2}(1+z)\ln z - (1-z), \quad (10)$$

$$P_{ee}^{(2,2)}(z) = \frac{1}{2}\left[P_{e\gamma}^{(1)} \otimes P_{\gamma e}^{(1)}\right](z)$$

$$\equiv (1+z)\ln z + \frac{1}{2}(1-z) + \frac{2}{3}\frac{1}{z}(1-z^3), \quad (11)$$

$$P_{ee,f}^{(2,3)}(z) = N_c(f)Q_f^2 \frac{1}{3} P_{ee}^{(1)}(z)\theta\left(1-z-\frac{2m_f}{E_e}\right) \quad (12)$$

denoting double-photon radiation, scattering of a fermion into a fermion by a collinear photon, and collinear fermion pair production. $\otimes$ denotes the Mellin convolution

$$A(x) \otimes B(x) = \int_0^1 dx_1 \int_0^1 dx_2 \delta(x - x_1 x_2) A(x_1) B(x_2), \quad (13)$$

$Q_f$ is the fermion charge, and $N_c(f) = 3$ for quarks, $N_c(f) = 1$ for leptons, respectively.

The soft-photon exponentiation is performed solving the non-singlet evolution equation in the range $z \to 1$ analytically (cf. e.g. [31]). Since the terms up to $\mathcal{O}(\alpha^2)$ were taken into account in eq. (9) already the corresponding contributions have to be subtracted. One obtains [24]:

$$\frac{d^2\sigma^{(>2,soft)}}{dxdy} = \int_0^1 dz P_{ee}^{(>2)}(z,Q^2) \left\{\theta(z-z_0)\mathcal{J}(x,y,z)\frac{d^2\sigma^{(0)}}{dxdy}\bigg|_{x=\hat{x},y=\hat{y},S=\hat{S}} - \frac{d^2\sigma^{(0)}}{dxdy}\right\}, \quad (14)$$

with

$$P_{ee}^{>2}(z,Q^2) = D_{NS}(z,Q^2) - \frac{\alpha}{2\pi}L_e\frac{2}{1-z}\left\{1 + \frac{\alpha}{2\pi}L_e\left[\frac{11}{6} + 2\ln(1-z)\right]\right\}, \quad (15)$$

$$D_{NS}(z,Q^2) = \zeta(1-z)^{\zeta-1}\frac{\exp\left[\frac{1}{2}\zeta\left(\frac{3}{2} - 2\gamma_E\right)\right]}{\Gamma(1+\zeta)}, \quad (16)$$

$$\zeta = -3\ln\left[1 - (\alpha/3\pi)L_e\right]. \quad (17)$$

Finally, the fermion conversion term in $\mathcal{O}(\alpha^2 L_e^2)$ reads [24]:

$$\frac{d^2\sigma^{(2,e^- \to e^+)}}{dx dy} = \int_{z_0}^{1} dz P(z,Q^2;e^- \to e^+)\mathcal{J}(x,y,z)\frac{d^2\sigma^{(0)}}{dx dy}\bigg|_{x=\hat{x},y=\hat{y},S=\hat{S}}, \quad (18)$$

with the conversion rate given by

$$P(z,Q^2;e^- \to e^+) = \left(\frac{\alpha}{2\pi}\right)^2 L_e^2 P_{ee}^{(2,2)}(z). \quad (19)$$

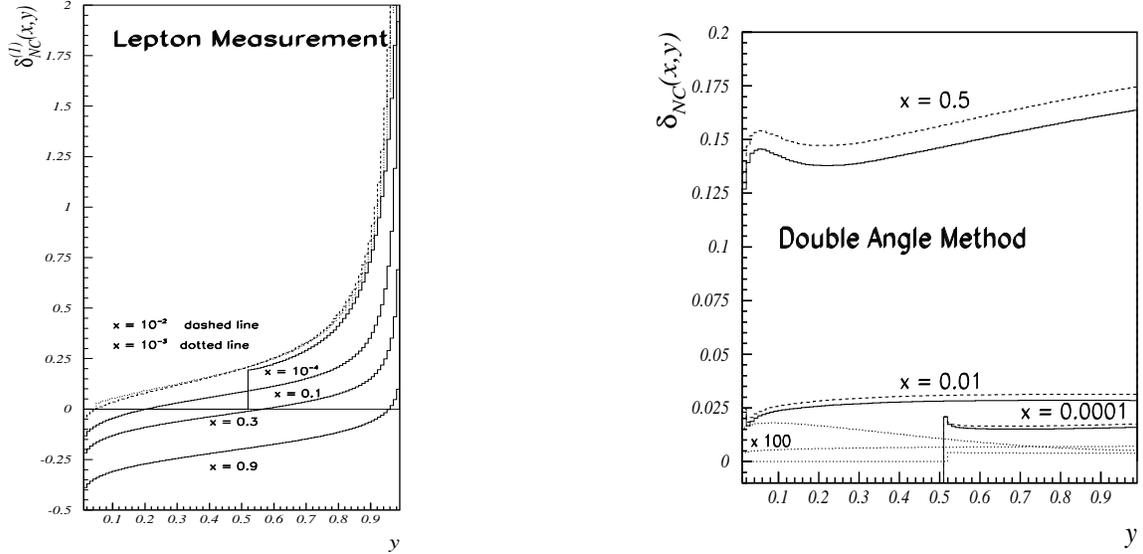

Figure 2: a) $\mathcal{O}(\alpha)$ initial state QED corrections to neutral current deep inelastic scattering for leptonic variables in LLA. b) initial state QED corrections using the double angle method. A z-cut of $E = 35\,\text{GeV}$ (cf. [24]) is applied. Full lines: LLA corrections up to second order + soft exponentiation; dashed lines: 1st order LLA corrections.

In figure 2 we illustrate the size of QED radiative corrections for two types of the measurement of the kinematical variables in the case of neutral current deep inelastic scattering. In the case of the (classical) leptonic variables $x$ and $Q^2$ are determined from the measurement of the scattered lepton only. In the double angle method the scattering angles of the outgoing lepton and hadronic jet are used on the other hand.[b] Whereas the radiative corrections for leptonic variables become very large at small $x$ and high $y$, they behave flat in $y$ in the case of the double angle method and are very small in the range $x \leq 0.01$, where the structure functions become rather large. Thus the latter set of variables behaves ideal in this respect.

The different methods to calculate the QED radiative corrections have been compared in $\mathcal{O}(\alpha)$ for a variety of kinematical measurements and are well understood. In 2nd order so far only LLA results are available [21,24] for the full set of outer kinematical variables studied by the HERA experiments.

## 3. The running coupling constant

The strong coupling $\alpha_s(\mu^2)$ is a central parameter in QCD. It is not an observable itself but the various hard scattering processes are often compared with respect to this quantity. Since it is scheme–dependent these comparisons have to be performed in a single renormalization scheme, as e.g. the $\overline{MS}$–scheme [32]. The running of $\alpha_s(\mu^2)$ is determined by the renormalization group equation [33].

$$\frac{d\alpha_s(\mu^2)}{\ln \mu^2} = -\frac{\beta_0}{4\pi}\alpha_s^2 - \frac{\beta_1}{(4\pi)^2}\alpha_s^3 - \frac{\beta_2}{(4\pi)^3}\alpha_s^4 + \ldots . \tag{20}$$

So far the contributions to the $\beta$-function have been calculated up to 3-loop order in the $\overline{MS}$–scheme, where the LO [34–36], NLO [37,38], and NNLO [39,40] terms are given by

$$\beta_0 = 11 - \frac{2}{3}N_f, \tag{21}$$

$$\beta_1 = 102 - \frac{38}{3}N_f, \tag{22}$$

$$\beta_2 = \frac{2857}{2} - \frac{5033}{18}N_f + \frac{325}{54}N_f^2. \tag{23}$$

$N_f$ denotes the number of active flavours. The solution of (20) reads:

$$\frac{1}{\alpha_s(Q^2)} = \frac{1}{\alpha_s(Q_0^2)} + \frac{\beta_0}{4\pi}\ln\left(\frac{Q^2}{Q_0^2}\right) + \Phi^{(n)}(\alpha_s(Q^2);\beta_i) - \Phi^{(n)}(\alpha_s(Q_0^2);\beta_i) \tag{24}$$

The superscript $n$ denotes the term at which the expansion of the $\beta$-function in (20) was truncated. In NNLO one obtains

$$\Phi^{(2)}(x;\beta_i) = -\frac{\beta_1}{8\pi\beta_0}\ln\left|\frac{16\pi^2 x^2}{16\pi^2\beta_0 + 4\beta_1\pi x + \beta_2 x^2}\right|$$

$$+\frac{\beta_1^2 - 2\beta_0\beta_2}{8\pi\beta_0\sqrt{4\beta_2\beta_0 - \beta_1^2}}\arctan\left(\frac{2\pi\beta_1 + \beta_2 x}{2\pi\sqrt{4\beta_0\beta_2 - \beta_1^2}}\right). \tag{25}$$

---

[b]For other types of measurements see [27] and references therein.

Note that

$$N_f \leq 5 : \quad 4\beta_0\beta_2 - \beta_1^2 > 0$$
$$N_f = 6 : \quad 4\beta_0\beta_2 - \beta_1^2 < 0. \tag{26}$$

Table 1:
Summary of recent measurements of $\alpha_s$ (from [42]) and results from lattice calculations (cf. [43]).

| Process | $\langle Q \rangle$ [GeV] | $\alpha_s(M_Z^0)$ | $\alpha_s(Q)$ exp. | $\alpha_s(Q)$ theor. | Theory |
|---|---|---|---|---|---|
| GLS (CCFR) | 1.73 | $0.107 {+0.007 \atop -0.009}$ | $+0.006 \atop -0.007$ | $+0.004 \atop -0.006$ | NNLO |
| $R_\tau$ (CLEO) | 1.78 | $0.116 \pm 0.003$ | 0.002 | 0.002 | NNLO |
| $R_\tau$ (ALEPH) | 1.78 | $0.122 \pm 0.003$ | 0.002 | 0.002 | NNLO |
| $R_\tau$ (OPAL) | 1.78 | $0.123 \pm 0.003$ | 0.002 | 0.002 | NNLO |
| $R_\tau$ (P. Raczka) | 1.78 | $0.120 \pm 0.003$ | 0.002 | 0.002 | NNLO |
| $\eta_c \to \gamma\gamma$ (CLEO) | 2.98 | $0.101 \pm 0.010$ | 0.008 | 0.006 | NLO |
| $\Upsilon(1S)$ (CLEO) | 9.46 | $0.111 \pm 0.006$ | 0.001 | 0.006 | NLO |
| $ep \to$ jets (H1) | 5–60 | $0.123 \pm 0.018$ | 0.014 | 0.010 | NLO |
| $pp \to W$jets (D0) | 80.6 | $0.121 \pm 0.014$ | 0.012 | 0.005 | NLO |
| DIS: | | | | | |
| $\nu F_2, xF_3$ | 5 | $0.111 \pm 0.006$ | 0.004 | 0.004 | NLO |
| $\mu F_2$ | 7.1 | $0.113 \pm 0.005$ | 0.003 | 0.004 | NLO |
| $e^+e^- \to$ jets (CLEO) | 10.53 | $0.113 \pm 0.006$ | 0.002 | 0.006 | NLO |
| $e^+e^- \to Z^0$: | | | | | |
| scal. viol. (ALEPH) | 91.2 | $0.127 \pm 0.011$ | - | - | NLO |
| ev. shapes (SLD) | 91.2 | $0.120 \pm 0.008$ | 0.003 | 0.008 | resum. |
| $\gamma(Z^0 \to$ had.) (LEP) | 91.2 | $0.127 \pm 0.006$ | 0.005 | $+0.003 \atop -0.004$ | NNLO |
| LGT | | 0.111 to 0.115 | | 0.005 to 0.007 | [43] |

One may solve eq. (20) using $\alpha_s(Q_0) \equiv \alpha_s(M_Z)$ as input. Due to the fact that $N_f = 5$ for $Q < 10$ GeV the NNLO correction diminishes the NLO solution slightly at low values of $Q$, cf. [41].

In table 1 recent measurements of $\alpha_s$ (cf. [42]) from different $Q$ ranges are compared, which were evaluated at the $Z^0$ scale. We also added the range of recent results from lattice calculations (LGT) (cf. [43]). The most precise measurements stem from the various high statistics deep inelastic scattering experiments at the one side, and high precision measurements of different observables in $e^+e^-$ annihilation, on the other side. The average values for these measurements are:

$$\text{DIS} : \overline{\alpha_s}(M_Z) = 0.112 \pm 0.004 \tag{27}$$
$$e^+e^- : \overline{\alpha_s}(M_Z) = 0.121 \pm 0.004, \tag{28}$$

showing a $2\sigma$ difference at present. The results from lattice calculations yield values of $\alpha_s(M_Z) = 0.111...0.115$. Here, the systematical error is estimated to be of $\mathcal{O}(0.005...0.007)$ still [43]. Whereas the pure gauge–field contributions are rather well understood the quark terms deserve more detailed investigations in the future to obtain decisive results.

## 4. The Evolution Equation

The structure functions, $F_j(x,Q^2)$, describing the deep inelastic scattering cross sections may be expressed as a convolution of the bare parton densities, $\hat{f}_i$, and hard scattering cross sections, $\sigma_j^i$,

$$F_j(x,Q^2) = \hat{f}_i(x) \otimes \sigma_j^i(\alpha_s, Q^2/\mu^2, x, \varepsilon). \tag{29}$$

The functions $\sigma_j^i$ contain initial state mass- and ultraviolet singularities. If the calculation is performed in $4-\varepsilon$ dimensions they emerge as poles in $\varepsilon$. The mass singularities can be factorized and absorbed into the bare parton densities [44]. The ultraviolet singularities are removed by the renormalization of the bare coupling constant, $\alpha_s$. In this way two scales, the mass factorization scale, $\mu_1$, and the renormalization scale, $\mu_2$, are introduced. One further separates $\sigma_j^i$ into its pole-, $\Gamma_i^k$, and non-pole parts, $C_{jk}$. This separation is obviously arbitrary and introduces a scheme-dependence. One obtains:

$$F_j(x,Q^2) = \hat{f}^i(x) \otimes \Gamma_i^k(\alpha_s(\mu_2^2), \mu_1^2/\mu^2, \mu_1^2/\mu_2^2, \varepsilon) \otimes C_{jk}(\alpha_s(\mu_2^2), Q^2/\mu_1^2, \mu_1^2/\mu_2^2, x). \tag{30}$$

The first two factors on the rhs of (30) define the renormalized parton densities, $f^i$, which are scheme-dependent. One may identify the scales $\mu_1 = \mu_2 = M$ and obtains

$$F_j(x,Q^2) = f^k(x, \alpha_s(M^2), M^2/\mu^2) \otimes C_{jk}(\alpha_s(M^2), Q^2/M^2, x). \tag{31}$$

Let us transform the above equations to moment space by

$$\int_0^1 dx\, x^{N-1} \Delta(x) \equiv \Delta(N). \tag{32}$$

The invariance of the structure functions $F_i$ against the choice of $M$ may be expressed by

$$\left[ M\frac{\partial}{\partial M} + \beta(g)\frac{\partial}{\partial g} - 2\gamma_\psi(g) \right] F_j(N) = 0 \tag{33}$$

leading to

$$\left[ M\frac{\partial}{\partial M} + \beta(g)\frac{\partial}{\partial g} + \gamma_\kappa^N(g) - 2\gamma_\psi(g) \right] f_k(N) = 0, \tag{34}$$

$$\left[ M\frac{\partial}{\partial M} + \beta(g)\frac{\partial}{\partial g} - \gamma_\kappa^N(g) \right] C_{jk}(N) = 0, \tag{35}$$

where we use a generic notation for the anomalous dimensions, $\gamma_\kappa^N$. Here, eq. (34) describes the evolution of the parton densities, which results *directly* from the renormalization group equation. The anomalous dimensions $\gamma_\kappa$ are the Mellin transforms of the splitting functions, which will be defined below.

Let us introduce some combinations of parton densities:

$$q_i^-(x, Q^2) = q_i(x, Q^2) - \bar{q}_i(x, Q^2) \tag{36}$$

$$q_i^+(x, Q^2) = q_i(x, Q^2) + \bar{q}_i(x, Q^2) \tag{37}$$

$$q^+(x, Q^2) = \sum_{i=1}^{N_f} q_i^+(x, Q^2) \tag{38}$$

$$\tilde{q}_i^-(x, Q^2) = q_i^+(x, Q^2) - \frac{1}{N_f} q^+(x, Q^2). \tag{39}$$

$q_i^-$ and $\tilde{q}_i^-$ denote flavour non-singlet combinations, and $q^+$ is the singlet density. The Mellin transform of (34) yields the non-singlet and singlet evolution equations in $x$ space:

$$\frac{d}{d \ln Q^2} q_i^-(x, Q^2) = \frac{\alpha_s(Q^2)}{2\pi} P^-(x, \alpha_s) \otimes q_i^-(x, Q^2) \tag{40}$$

$$\frac{d}{d \ln Q^2} \tilde{q}_i^-(x, Q^2) = \frac{\alpha_s(Q^2)}{2\pi} P^+(x, \alpha_s) \otimes \tilde{q}_i^-(x, Q^2) \tag{41}$$

$$\frac{d}{d \ln Q^2} \begin{bmatrix} q^+(x, Q^2) \\ G(x, Q^2) \end{bmatrix} = \frac{\alpha_s(Q^2)}{2\pi} \boldsymbol{P}(x, \alpha_s) \otimes \begin{bmatrix} q^+(x, Q^2) \\ G(x, Q^2) \end{bmatrix}. \tag{42}$$

The splitting functions, $P^\pm$ and $\boldsymbol{P}$ can be expressed by the following perturbative series:

$$P^\pm(x, \alpha_s) = P_{NS}^{(0)} + \frac{\alpha_s}{2\pi} P^{\pm,(1)}(x) + \left(\frac{\alpha_s}{2\pi}\right)^2 P^{\pm,(2)}(x) + ... \tag{43}$$

$$\boldsymbol{P}(x, \alpha_s) = \boldsymbol{P}^{(0)} + \frac{\alpha_s}{2\pi} \boldsymbol{P}^{(1)}(x) + \left(\frac{\alpha_s}{2\pi}\right)^2 \boldsymbol{P}^{(2)}(x) + ... \tag{44}$$

Let us change the evolution scale $Q \equiv M$ introducing

$$t := -\frac{2}{\beta_0} \ln \frac{\alpha_s(Q^2)}{\alpha_s(Q_0^2)} \tag{45}$$

$$\frac{\alpha_s(Q^2)}{2\pi} d \ln Q^2 = \left(1 - \frac{\beta_1}{2\beta_0} \frac{\alpha_s(Q^2)}{2\pi} + ...\right) dt. \tag{46}$$

Then, the evolution equations may be rewritten as:

$$q_i^-(x, t) := E^-(x, t) \otimes q_i^-(x) \tag{47}$$

$$q_i^+(x, t) := E^+(x, t) \otimes q_i^+(x, t) + \frac{1}{N_f} \left[E_{11}(x, t) - E^+(x, t)\right] \otimes q^+(x)$$

$$+\frac{1}{N_f}E_{12}(x,t) \otimes G(x) \tag{48}$$

$$\begin{bmatrix} q_i^+(x,t) \\ G(x,t) \end{bmatrix} = \boldsymbol{E}(x,t) \otimes \begin{bmatrix} q_i^+(x,t) \\ G(x,t) \end{bmatrix}, \tag{49}$$

where we introduced the evolution operators $E^\pm$ and $\boldsymbol{E}$ obeying the initial conditions

$$\lim_{t \to 0} E^\pm(x,t) = \delta(1-x) \tag{50}$$

$$\lim_{t \to 0} \boldsymbol{E}^\pm(x,t) = \boldsymbol{1}\delta(1-x). \tag{51}$$

They allow to *separate* the non-perturbative input densities $q_i^\pm(x), q^+(x)$ and $G(x)$ from those terms which can be calculated perturbatively. The evolution operators itself obey the evolution equations:

$$\frac{d}{dt}E^\pm(x,t) = \left\{P_{NS}(x) + \frac{\alpha_s(t)}{2\pi}R^\pm(x) + ...\right\} \otimes E^\pm(x,t) \tag{52}$$

$$\frac{d}{dt}\boldsymbol{E}(x,t) = \left\{\boldsymbol{P}^{(0)}(x) + \frac{\alpha_s(t)}{2\pi}\boldsymbol{R}(x) + ...\right\} \otimes \boldsymbol{E}(x,t), \tag{53}$$

where

$$R^\pm(x) = P^{\pm,(1)}(x) - \frac{\beta_1}{2\beta_0}P_{NS}^{(0)} \tag{54}$$

$$\boldsymbol{R}(x) = \boldsymbol{P}^{(1)}(x) - \frac{\beta_1}{2\beta_0}\boldsymbol{P}^{(0)}. \tag{55}$$

*4.1. Splitting Functions*

The well-known leading order singlet splitting functions are given by[45-50]:

$$P_{qq}^{(0)}(z) = C_F\left[\frac{1+z^2}{(1-z)_+} + \frac{3}{2}\delta(1-z)\right] \tag{56}$$

$$P_{qg}^{(0)}(z) = T_f N_f\left[z^2 + (1-z)^2\right] \tag{57}$$

$$P_{gq}^{(0)}(z) = C_F\frac{1+(1-z)^2}{z} \tag{58}$$

$$P_{gg}^{(0)}(z) = 2C_A\left[\frac{1-z}{z} + \frac{z}{(1-z)_+}\right] + \frac{1}{2}\beta_0\delta(1-z), \tag{59}$$

and the non–singlet splitting funtion obeys $P_{NS}^{(0)}(z) \equiv P_{qq}^{(0)}(z)$. Here, $C_A = N_c = 3$, $C_F = (N_c^2 - 1)/N_c$, and $T_f = 1/2$. In leading order the splitting functions for space and timelike virtualities are the same [31]. This relation is violated in higher orders.

The non-singlet splitting functions in NLO[51−55] for spacelike virtualities are given by:

$$P^{\pm}(z, \alpha_s) = \widehat{P}^{\pm}(z, \alpha_s) - \delta(1-z) \int_0^1 dz \widehat{P}^{-}(z, \alpha_s), \tag{60}$$

$$\widehat{P}^{\pm}(z, \alpha_s) = \widehat{P}_{qq}(z, \alpha_s) \pm \widehat{P}_{q\bar{q}}(z, \alpha_s), \tag{61}$$

where the superscript $\pm$ labels the type of non-singlet evolution, see eqs. (40, 41).

$$\widehat{P}_{qq}(z, \alpha_s) = \left(\frac{\alpha_s}{2\pi}\right) C_F \left(\frac{1+z^2}{1-z}\right)$$
$$+ \left(\frac{\alpha_s}{2\pi}\right)^2 \left[C_F^2 P_F(z) + \frac{1}{2} C_F C_A P_G(z) + C_F N_f T_f P_{N_f}(z)\right] \tag{62}$$

$$\widehat{P}_{q\bar{q}}(z, \alpha_s) = \left(\frac{\alpha_s}{2\pi}\right)^2 \left[C_F^2 - \frac{1}{2} C_F C_A\right] P_A(z) \tag{63}$$

$$P_F(z) = -2\frac{1+z^2}{1-z} \ln z \ln(1-z) - \left(\frac{3}{1-z} + 2z\right) \ln z - \frac{1}{2}(1+z)\ln^2 z$$
$$-5(1-z) \tag{64}$$

$$P_G(z) = \frac{1+z^2}{1-z}\left[\ln^2 z + \frac{11}{3}\ln z + \frac{67}{9} - \frac{1}{3}\pi^2\right] + 2(1+z)\ln z + \frac{40}{3}(1-z) \tag{65}$$

$$P_{N_f}(z) = -\frac{2}{3}\left[\frac{1+z^2}{1-z}\left(\ln z + \frac{5}{3}\right) + 2(1-z)\right] \tag{66}$$

$$P_A(z) = 2\frac{1+z^2}{1-z}\int_{z/(1+z)}^{1/(1+z)} \frac{du}{u} \ln\left(\frac{1-u}{u}\right) + 2(1+z)\ln z + 4(1-z) \tag{67}$$

The NLO singlet splitting functions were derived in refs. [55−58] for the unpolarized case both for space- and timelike virtualities.

### 4.2. Coefficient Functions

In the calculation of the higher order corrections to the structure functions the coefficient functions are required (cf. eq. (31)). These are scheme-dependent quantities. In the $\overline{MS}$-scheme they read in $\mathcal{O}(\alpha_s)$ [59]:

$$C_{F2}^{(1)}(z) = C_F\left[\frac{1+z^2}{1-z}\left(\ln\frac{1-z}{z} - \frac{3}{4}\right) + \frac{1}{4}(9+5z)\right]_+ \tag{68}$$

$$C_{F1}^{(1)}(z) = C_{F2}^{(1)}(z) - 2zC_F \tag{69}$$

$$C_{F3}^{(1)}(z) = C_{F2}^{(1)}(z) - C_F(1+z) \tag{70}$$

$$C_{G2}^{(1)}(z) = 2N_f T_f \left\{\left[z^2 + (1-z)^2\right]\ln\frac{1-z}{z} - 1 + 8z(1-z)\right\} \tag{71}$$

$$C_{G1}^{(1)}(z) = C_{G2}^{(1)}(z) - 8N_f T_f z(1-z). \tag{72}$$

The $\mathcal{O}(\alpha_s^2)$ contributions to the coefficient functions were calculated in refs. [60-63] for the structure functions $F_2$, $F_L$, and $xF_3$. Whereas in refs. [61-63] the coefficient functions were derived in $z$-space, the moments $M^n|_{n=2,...10}$ were calculated for $F_L$ and $F_2$ in ref. [60]. The results of both calculations do fully agree.

To illustrate the numerical importance of the $\mathcal{O}(\alpha_s^2)$ calculation we compare in figure 3 the $\mathcal{O}(\alpha_s)$ and the $\mathcal{O}(\alpha_s^2)$ result for the structure function $F_L(x, Q^2)$ in the $\overline{\text{MS}}$–scheme using the parametrization [64] for the parton densities.

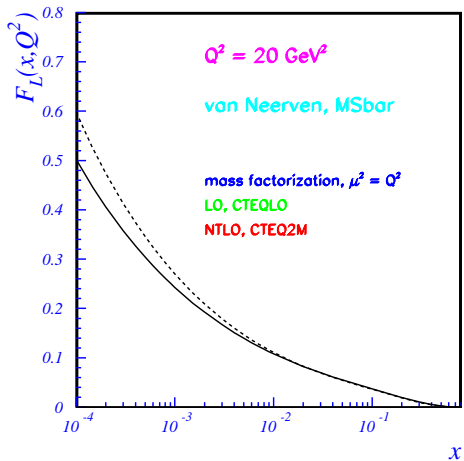

Figure 3: Comparison of QCD corrections to the structure function $F_L(x, Q^2)$. $\mathcal{O}(\alpha_s)$ (dashed line); $\mathcal{O}(\alpha_s^2)$ (full line); from [65].

The NLO correction leads to a depletion relatively to the LO result by $\sim 15\%$ at $x \sim 10^{-4}$. In the small $x$ range $F_L(x, Q^2)$ is widely determined by the gluon density. The $\mathcal{O}(\alpha_s^2)$ corrections are not negligible and have to be taken into account for the unfolding of the gluon density from a measurement of $F_L(x, Q^2)$.

4.3. $O(\alpha_s^3)$ corrections

For a series of observables related to deep inelastic scattering QCD corrections up to 3–loop order were calculated. The quantities which were studied so far are moments of structure functions and specific combinations which are related to sum–rules [66-68] of structure functions.

The $\mathcal{O}(\alpha_s^3)$ corrections to the Bjorken-[69], the Gross–Llewellyn Smith-[70], and the polarized Bjorken[71] sum rules are:

$$\int_0^1 dx \left[ F_1^{\bar{\nu}p}(x, Q^2) - F_1^{\nu p}(x, Q^2) \right] = 1 - \frac{2}{3}\frac{\alpha_s}{\pi} - 2.3519 \left(\frac{\alpha_s}{\pi}\right)^2 - 8.4852 \left(\frac{\alpha_s}{\pi}\right)^3 + ... \quad (73)$$

$$\int_0^1 dx \left[ F_3^{\bar{\nu}p}(x, Q^2) - F_3^{\nu p}(x, Q^2) \right] = 6 \left\{ 1 - \frac{\alpha_s}{\pi} + \left(\frac{\alpha_s}{\pi}\right)^2 \left(-\frac{55}{12} + \frac{1}{3}N_f\right) \right.$$

$$
\begin{aligned}
&+ \left(\frac{\alpha_s}{\pi}\right)^3 \left[-\frac{13841}{216} - \frac{44}{9}\zeta_3 + \frac{55}{2}\zeta_5 \right. \\
&\left. + N_f \left(\frac{10009}{1296} + \frac{91}{54}\zeta_3 - \frac{5}{3}\zeta_5\right) - \frac{115}{648}N_f^2 \right]\Big\} \quad (74)
\end{aligned}
$$

$$
\begin{aligned}
\int_0^1 dx \left[g_1^{ep}(x,Q^2) - g_1^{en}(x,Q^2)\right] &= \frac{1}{3}\left|\frac{g_A}{g_V}\right|\Big\{1 - \frac{\alpha_s}{\pi}... \\
&+ \left(\frac{\alpha_s}{\pi}\right)^3 \left[...N_f\left(\frac{10339}{1296} + \frac{61}{54}\zeta_3 - ...\right)...\right]\Big\} (75)
\end{aligned}
$$

Note the small difference in the correction factors between the Gross–Llewellyn-Smith and the polarized Bjorken sum rules in the $N_f$ term in $\mathcal{O}(\alpha_s^3)$.

These calculations are performed by fast formula manipulation programs as FORM [72]. Due to the complexity of the problem they request hundred(s) of CPU hours on present day computers. The non-singlet moments $M^n|_{n=2,4,6,8}$ of the structure functions $F_2(x,Q^2)$ and $F_L(x,Q^2)$ were also calculated [73]. The calculation of still higher moments for the non-singlet case and the first singlet moments of these structure functions is being performed currently [74].

Precise experimental data for the sum-rules and moments quoted will allow *very* concise tests of QCD in its perturbative range.

## 5. Resummation of small $x$ contributions

At small values of $x$ contributions to the splitting functions show a singular behaviour. This is both the case for the singlet and non-singlet splitting functions. Resummations of large terms are also required at large $x$ [75]. Here we will deal with the small $x$ behaviour only.

### 5.1. Singlet terms

In the singlet case the singularity in the splitting function is of the type $P \propto (1/x)\alpha_s^l \ln^{l-1} x$. In leading order these contributions are resummed by the BFKL equation [76][c]. The leading singular terms in the gluon anomalous dimension, $\gamma_{gg}$, to all orders in $\alpha_s$ are given by the solution of

$$
\begin{aligned}
N - 1 &= \overline{\alpha}_s \chi\left[\gamma_L(N, \overline{\alpha}_s)\right], && (76) \\
\chi(z) &= 2\psi(1) - \psi(z) - \psi(1-z), && (77)
\end{aligned}
$$

with $\overline{\alpha}_s = C_A \alpha_s/\pi$. The solution of (76) is multivalued and one has to select a single Riemann–sheet. This is done imposing the condition

$$
\lim_{|N|\to\infty} \gamma_L(N, \overline{\alpha}_s) \propto \frac{\overline{\alpha}_s}{N-1}, \quad N \in \mathbf{C}. \quad (78)
$$

---
[c]For a recent review see ref. [77].

In figure 4 real and imaginary part of this solution are shown. Asymptotically, i.e. for $A \equiv \overline{\alpha}_s/(N-1) \ll 1$, one obtains:

$$\gamma_L(N,\overline{\alpha}_s) = \frac{\overline{\alpha}_s}{N-1}\left\{1 + 2\sum_{k=1}^{\infty}\zeta_{2k+1}\gamma_L^{2k+1}(N,\overline{\alpha}_s)\right\}$$
$$\equiv A + 2\zeta_3 A^4 + 2\zeta_5 A^6 + 12\zeta_3^2 A^7 + \ldots. \qquad (79)$$

The solution of (76) using (78) contains three branch points. They are solutions of the equation ($z \equiv \gamma_L^s$):

$$\psi'(z) - \frac{\pi^2}{2}\frac{1}{\sin^2 \pi z} = 0, \qquad (80)$$

and are [78,79] given by:

$$\gamma_{L1}^s = 1/2 \qquad (81)$$
$$\gamma_{L2,3}^s = -0.4252 \pm 0.4739i . \qquad (82)$$

The former value is seen in figure 4 as the 'roof' in Re$\gamma$, the latter ones are the edges at $\rho = -1.4105 \pm 1.9721i$.

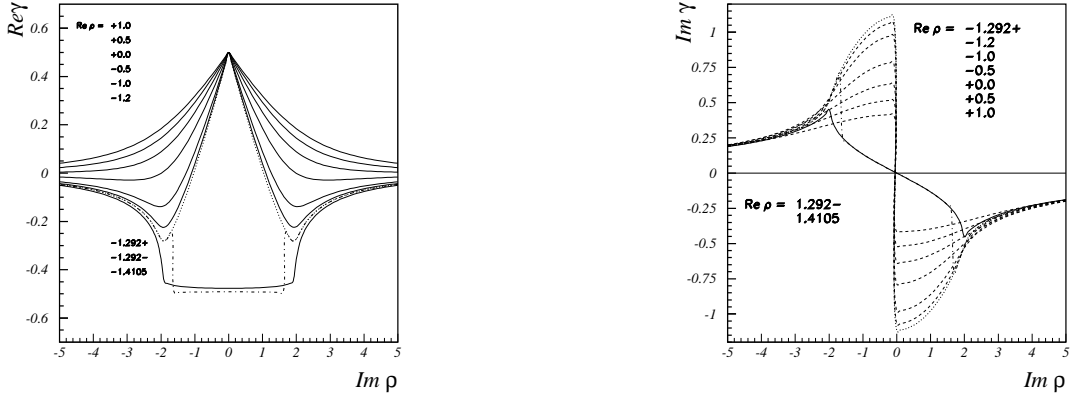

Figure 4: The solution of eq. (76) for complex values of $\rho \equiv (N-1)/\overline{\alpha}_s$. a) Re($\gamma_L$); b) Im($\gamma_L$), cf. [79].

The resummation for the most singular terms in the singlet anomalous dimension takes the form [80] :

$$\gamma_{ab}(N,\alpha_s) = \sum_{k=1}^{\infty}\left(\frac{\alpha_s}{N-1}\right)^k A_{ab}^{(k)} + \sum_{k=1}^{\infty}\alpha_s\left(\frac{\alpha_s}{N-1}\right)^k B_{ab}^{(k)} + O\left(\alpha_s^2\left(\frac{\alpha_s}{N-1}\right)^k\right). \qquad (83)$$

The LO and NLO terms to this matrix equation are given by

$$\mathbf{\Gamma}_L(N) = \begin{pmatrix} 0 & 0 \\ \frac{C_F}{C_A}\gamma_L(N) & \gamma_L(N) \end{pmatrix} \qquad (84)$$

$$\mathbf{\Gamma}_{NL}(N) = \begin{pmatrix} \frac{C_F}{C_A}\gamma_{NL}(N) - \frac{2\alpha_s}{3\pi}T_f & \gamma_{NL}(N) \\ \gamma_\delta(N) & \gamma_\eta(N) \end{pmatrix}. \qquad (85)$$

So far only the quark contributions in the NLO terms are calculated. These, and the LO terms, are functions of $\gamma_L$. An approximate solution for $\gamma_{NL}$ reads [80][d]

$$\gamma_{NL}(N) \simeq \frac{2\alpha_s}{3\pi}T_f\left\{1 + 2.17\frac{\alpha_s}{N-1} + 2.30\left(\frac{\alpha_s}{N-1}\right)^2 + 8.27\left(\frac{\alpha_s}{N-1}\right)^3 + ...\right\}. \qquad (86)$$

One may use the expression (83), supplemented by the terms contributing to $(\gamma_{ab})$ up to NLO (cf. section 4), and solve the singlet evolution equation. This was done in [83,78] assuming a flat input at $Q_0^2 = 4\,\text{GeV}^2$.

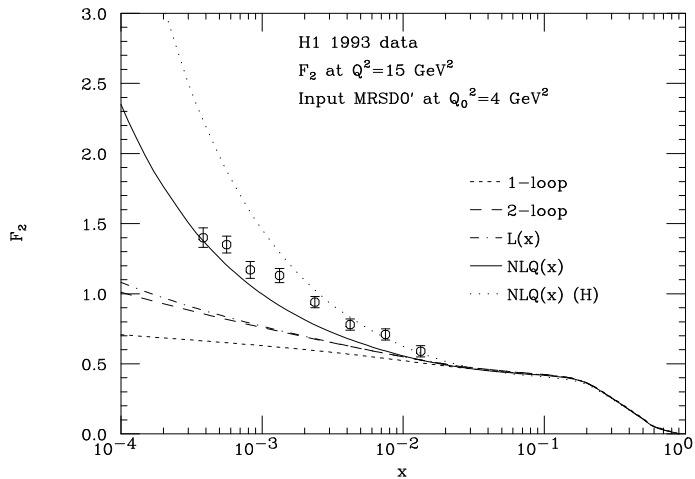

Figure 5: Resummed predictions for the structure function $F_2(x, Q^2)$ [78], see text.

In figure 5 the behaviour of $F_2(x, Q^2)$ due to different contributions to the evolution is illustrated. The evolution of the flat input using the pure LO and NLO terms leads to a small rise at low $x$. The resummed LO terms affect $F_2$ through the gluon density only and lead to a correction which is rather small if compared to the correction due to the resummed NLO terms, which affect the quark densities directly. Here, for the yet unknown entries $\gamma_{\delta,\eta}$ in eq. (85) the NLO terms were used. Since the resummation (83) is performed for the most singular terms at small $x$ only, momentum conservation has to be restored explicitly. This is possible in different ways. One may either introduce an appropriate term $\propto \delta(1-z)$ in the splitting functions or multiply the anomalous dimensions by $(1-N)$. The latter solution is illustrated by the full line in figure 5, while the former is shown as the dotted line. The large difference between the two solutions shows that it is likely that sub-dominant terms may be as important for the scaling violations of $F_2$ as the singular terms at small $x$.

---

[d]Related numerical studies were performed in ref. [81,82] also.

A unified form for a leading order evolution eqution accounting both for the small $x$ terms due to the BFKL equation and the evolution kernels [45–50] was found in [84–86]. It is based on the angular ordering of the gluon cascade. In the limit of small $x$ the BFKL equation is obtained, while for medium and large values of $x$ this equation turns into the LO evolution equation of pure gluodynamics, i.e. the angular ordering turns into strong ordering of $k_\perp$. This concept was worked out in leading order so far. A numerical illustration of this resummation was given in [87].

*5.2. Non-singlet terms*

As shown in sect. 4.1 the most singular terms in the non-singlet splitting functions in $\mathcal{O}(\alpha_s)$ and $\mathcal{O}(\alpha_s^2)$ behave as $\alpha_s(\alpha_s \ln^2 x)^k$. In ref. [88] a resummation of these contributions to the $\pm$ combinations (cf. eq. (60)) of non-singlet structure functions was derived. Recently very sizeable corrections [89] due to this resummation have been claimed both for unpolarized and polarized structure functions.

Similar to the considerations in section 5.1 the resummation can be studied in the context of the renormalization group equation, see ref. [90]. Unlike the case of the BFKL equation the present resummation deals *not* with the anomalous dimension but with the structure function *itself*. Therefore one has to consider the evolution equation for the non-singlet structure functions here:

$$\frac{\partial F^\pm_{NS,i}(x, a_s)}{\partial a_s} = -\frac{1}{\beta_0 a_s^2} K^\pm_i(x, a_s) \otimes F^\pm_{NS,i}(x, a_s), \qquad (87)$$

where $a_s = \alpha_s(Q^2)/(4\pi)$. In NLO the evolution kernels $K^\pm_{NS,1}$ are

$$K^\pm_{i,1}(x, a_s) = P_{NS,0}(x) a_s + \left[ P^\pm_{NS,1}(x) - \frac{\beta_1}{\beta_0} P_{NS,0}(x) - \beta_0 c^\pm_{i,1}(x) \right] a_s^2 \qquad (88)$$

The labels $\pm$ denote the type of the non-singlet evolution. The combination $F_2^{ep} - F_2^{en}$, e.g., belongs to the '+' type, and $xF_3^{\nu N} + xF_3^{\bar{\nu} N}$ and $g_1^{ep} - g_1^{en}$ are '−' type combinations. In the latter case the splitting functions obey $\int_0^1 dx P_l^-(x) = 0$ in each order in $\alpha_s$ due to fermion number conservation.

The resummation of the most singular parts in the kernels $K^\pm_i(x, a_s)$ read in $N$-space [88]:

$$\Gamma^+_{NS, x\to 0}(N) = -2N \left\{ 1 - \sqrt{1 - \frac{2\alpha_s C_F}{\pi N^2}} \right\} \qquad (89)$$

$$\Gamma^-_{NS, x\to 0}(N) = -2N \left\{ 1 - \sqrt{1 - \frac{2\alpha_s C_F}{\pi N^2} \left[ 1 - \frac{2N_c \alpha_s}{\pi N} \frac{d}{dN} \ln \left( e^{z^2/4} D_{-1/2N_c^2}(z) \right) \right]} \right\}, \qquad (90)$$

where $N = z\sqrt{\overline{\alpha_s}/2}$, and $\overline{\alpha_s} = N_c \alpha_s/\pi$. $D_p(x)$ denotes the function of the parabolic cylinder. Up to $\mathcal{O}(\alpha_s^2)$ the coefficient functions in the $\overline{MS}$-scheme behave at most $\propto \ln^{2k-1} x$. Therefore the contributions to $K^{\pm}_{NS, x \to 0}(x, a_s)$ can be directly compared with the results from fixed order perturbation theory at least up to NNLO in the $\overline{\text{MS}}$-scheme. These are known up to NLO and are found to agree [90].

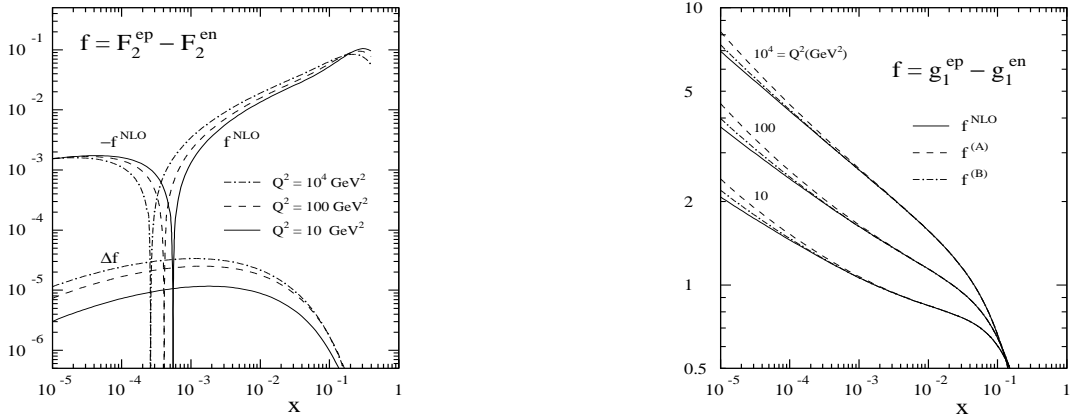

Figure 6: The small-$x$ $Q^2$ evolution of non-singlet structure functions in NLO and the correction due to the resummed kernels eqs. (89,90)[90]: a) $F_2^{ep} - F_2^{en}$; b) $g_1^{ep} - g_1^{en}$. The labels $A$ and $B$ refer to the way in which fermion number conservation is restored. A: $\Gamma^-(N, a_s) \to \Gamma^-(N, a_s) - \Gamma^-(1, a_s)$, B: $\Gamma^-(N, a_s) \to \Gamma^-(N, a_s) \cdot (1 - N)$.

In figure 6 we compare the evolution for $+$ and $-$ non-singlet combinations of structure functions in NLO with the resummed contributions beyond NLO. The latter terms yield corrections at the level of $\mathcal{O}(1\%)$ in the accessible kinematical ranges (cf. [91]). Moreover, in the case of $-$ type combinations different possible ways to restore fermion number conservation lead to a variation of the terms beyond NLO by a factor of three. This signals that yet unknown medium-$x$ contributions to the evolution kernels may be as important as the small $x$ terms. A similar behaviour was observed in the case of the singlet terms in sect. 5.1.

## 6. Heavy flavour contributions to structure functions

In lowest order the contributions to the heavy flavour structure functions $F_{2,L}(x, Q^2)$ for the $|\gamma|^2$ term in neutral current deep inelastic scattering are described by the diagrams due to photon–gluon fusion. They are given by

$$F_{2,L}^{Q\overline{Q}}(x, Q^2, m_Q^2) = 2e_Q^2 x \frac{\alpha_s(\mu^2)}{2\pi} \int_{ax}^{1} \frac{dy}{y} C_{g;2,L}^{Q}\left(\frac{x}{y}, \frac{m_Q^2}{Q^2}\right) G(y, \mu^2), \tag{91}$$

where

$$\begin{aligned} C_{g,2}^{Q}\left(z, \frac{M_Q^2}{Q^2}\right) &= \frac{1}{2}\left\{\left[z^2 + (1-z)^2 + z(1-3z)\frac{4m_Q^2}{Q^2} - z^2\frac{8m_Q^2}{Q^4}\right] \ln \frac{1+\beta}{1-\beta} \right. \\ &\quad \left. + \beta\left[-1 + 8z(1-z) - z(1-z)\frac{4m_Q^2}{Q^2}\right]\right\}, \end{aligned} \tag{92}$$

$$C_{g,L}^{Q}\left(z, \frac{M_Q^2}{Q^2}\right) = -z^2 \frac{4m_Q^2}{Q^2} \ln\frac{1+\beta}{1-\beta} + 2\beta z(1-z), \tag{93}$$

with $a = 1 + 4M_Q^2/Q^2$ and $\beta^2 = 1 - (4m_Q^2/Q^2)z(1-z)^{-1}$. These contributions and the other LO terms were derived in refs. [92–99].

The NLO contributions were calculated in ref. [100–103]. Phenomenological studies can be found in [104,105]. The NLO corrections stabilize the numerical values of the scattering cross section with respect to the choice of the factorization scale.

In figure 7 numerical results are shown for the LO and NLO contributions to $F_2^{c\bar{c}}(x, Q^2)$ [101] illustrating the size of the NLO corrections vs the LO term.

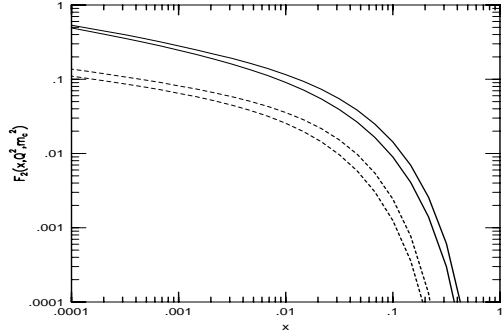

Figure 7: The $x$-dependence of $F_2^{LO,c\bar{c}}(x, Q^2, m_c^2)$ (lower pair) and $F_2^{NLO,c\bar{c}}(x, Q^2, m_c^2)$ (upper pair) at fixed $Q^2$. The solid lines are for $Q^2 = 100$ GeV$^2$ and the dashed lines are for $Q^2 = 10$ GeV$^2$; (from [101]).

## 7. $J/\psi$ production

As in the case of heavy flavour production from the measurement of the deep inelastic production cross section of $J/\psi$ particles the gluon density can be determined. In lowest order the cross section for the photoproduction case is given by [106]

$$\frac{d\sigma^0}{dt_1} = \frac{128\pi^2}{3} \frac{\alpha \alpha_s^2 e_c^2}{s^2} M_{J/\psi}^2 \frac{|\Phi(0)|^2}{M_{J/\psi}} \frac{s^2 s_1^2 + t^2 t_1^2 + u^2 u_1^2}{s_1^2 t_1^2 u_1^2} \tag{94}$$

in the colour-singlet model. Here we used the abbrevation $r_i = r - M_{J/\psi}^2, r \equiv s, t, u$. The scattering cross section for finite photon virtualities ($Q^2 > 0$) was derived in [107,108]. At a photon beam energy of $E_\gamma = 150$ GeV, e.g., the production cross section is lowered by a factor of $\sim 7$ for $Q^2 = 20$ GeV$^2$ compared to $Q^2 = 0$.

Recently the NLO corrections to the photoproduction cross section have been calculated [109,110]. A stabilization of the scale behaviour in the range $Q^2/m_c^2 > 1.5$ was obtained in comparison with the leading order result.

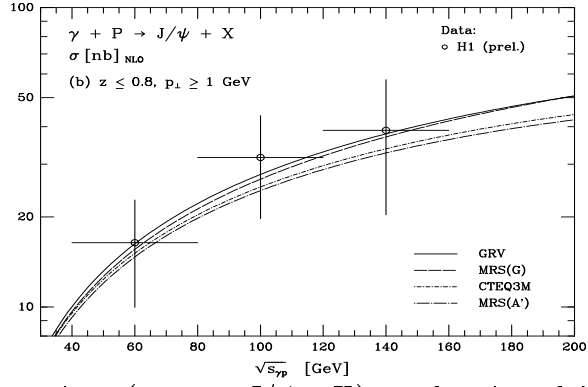

Figure 8: The total cross section $\sigma(\gamma + p \to J/\psi + X)$ as a function of the photon-proton CMS energy for different parametrizations of the parton densities (from [110]).

In figure 8 the total production cross section (NLO) is shown and compared with recent measurements at HERA.

## 8. QCD corrections to polarized structure functions

The leading order singlet splitting functions for polarized deep inelastic scattering [111,112,48] are :

$$P_{qq}^{(0)}(z) = C_F \left[ \frac{1+z^2}{(1-z)_+} + \frac{3}{2}\delta(1-z) \right] \qquad (95)$$

$$P_{qg}^{(0)}(z) = T_f N_f \left[ z^2 - (1-z)^2 \right] \qquad (96)$$

$$P_{gq}^{(0)}(z) = C_F \frac{1-(1-z)^2}{z} \qquad (97)$$

$$P_{gg}^{(0)}(z) = C_A \left[ \frac{2}{(1-z)_+} - 4z + 2 \right] + \frac{\beta_0}{2}\delta(1-z) \qquad (98)$$

Again the non singlet splitting function obeys $P_{NS}(x) \equiv P_{qq}(x)$. The non-singlet splitting functions in NLO are known from the unpolarized case (see sect. 4.1) already. Recently also the singlet splitting functions in NLO were calculated [113] in the $\overline{MS}$-scheme. The result of this calculation has been confirmed in [114] recently. It is interesting to note that eqs. (96, 97) differ from eqs. (57, 58) by relative signs in some of the terms. A common characteristics of these quantities is their leading singularity behaviour at small $x$. Unlike the unpolarized case the Mellin transform of the leading terms behave $\sim 1/N^k$.

The LO coefficient functions in the $\overline{MS}$-scheme are[115-117]

$$C_{q,NS}(z) = \delta(1-z) + \frac{\alpha_s}{4\pi} C_F \left\{ 4\left(\frac{\ln(1-z)}{1-z}\right)_+ - 3\left(\frac{1}{1-z}\right)_+ - 2(1+z)\ln(1-z) \right.$$
$$\left. - 2\frac{1+z^2}{1-z}\ln(z) + 4 + 2z - \delta(1-z)(4\zeta(2) + 9) \right\}, \qquad (99)$$

$$C_g(z) = \frac{\alpha_s}{4\pi} N_f T_f \left\{ 4(2z-1)\left[\ln(1-z) - \ln(z)\right] + 4(3-4z) \right\}. \qquad (100)$$

The NLO coefficient functions were calculated in ref. [118].

With these quantities at hand the scaling violations of the structure function $g_1(x, Q^2)$ can be studied up to NLO. This has been done recently[119,120] extending earlier studies in leading order. Still the parametrizations of the polarized parton densities do widely vary. Particularly this holds for extrapolations to the small $x$ and large $Q^2$ ranges.

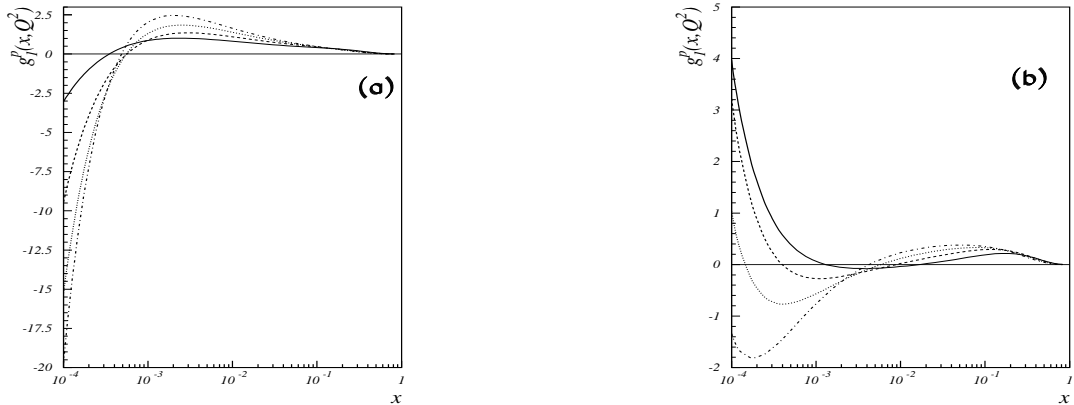

Figure 9: LO Parametrizations of the structure function $g_1^p(x, Q^2)$ in the range $x > 10^{-4}$. Full line: $Q^2 = 10 \, \text{GeV}^2$, dashed line: $Q^2 = 10^2 \, \text{GeV}^2$, dotted line: $Q^2 = 10^3 \, \text{GeV}^2$, dash–dotted line: $Q^2 = 10^4 \, \text{GeV}^2$ (cf. [91]); a) parametrization ref. [121]; b) parametrization ref. [122].

A systematic collection of the parametrizations for polarized parton densities and their evolution in LO (later than 1989) can be found in [123].

## 9. Open Problems

At the end of this survey I would like to list a series of open problems, the solution of which is of importance for the forthcoming development of the understanding of deep inelastic scattering in the perturbative range.

Concerning the QED corrections $\mathcal{O}(\alpha^2 L_e)$ terms should be calculated for sets of variables in which the 2nd order LLA terms are still large. Furthermore, complete $\mathcal{O}(\alpha)$ calculations should be performed for some sets of kinematical variables for which only LLA results exist to further improve the numerical accuracy.

The knowledge of the 3–loop non-singlet and singlet splitting functions for unpolarized deep inelastic scattering would be important to perform QCD tests at the level of NNLO corrections using measurements of the structure functions $F_2(x, Q^2)$, $W_2(x, Q^2)$, and $xW_3(x, Q^2)$. As a by-product of these calculations also a partial test of the validity of the resummations discussed in section 5 would be obtained.

The calculation of the NLO corrections to the BFKL equation are needed for futher studies of the behaviour of the anomalous dimension at small $x$.

NLO corrections should be calculated for different processes in polarized lepton-polarized nucleon scattering. They are also not yet performed for deep inelastic leptoproduction of $J/\psi$ particles at $Q^2 > 0$.

Besides of the twist-2 contributions to structure functions being discussed in the present paper the understanding of higher twist terms is important. Particularly gluonic twist-4 contributions to $F_2(x, Q^2)$ would be interesting to be derived in a complete calculation to compare with results obtained in Regge approaches for the limit of small $x$. These terms may play an important role in understanding the screening of the gluon density.

**Acknowledgements** I would like to thank W.L. van Neerven, S. Riemersma, and A. Vogt for interesting discussions.